\documentclass[11pt]{article}
\usepackage[vmargin=1in,hmargin=1in]{geometry}
\usepackage[english]{babel}
\usepackage[utf8]{inputenc}
\usepackage{amsmath,amssymb}
\usepackage{graphicx}
\usepackage{xcolor}
\usepackage{authblk}
\usepackage{floatrow}
\newfloatcommand{capbtabbox}{table}[][\FBwidth]

\usepackage{setspace}
\usepackage{hyperref}
\hypersetup{colorlinks=true}

\def\x{{\mathbf x}}
\def\s{{\mathbf s}}
\def\w{{\mathbf w}}
\def\y{{\mathbf y}}
\def\z{{\mathbf z}}
\def\W{{\mathbf W}}
\def\M{{\mathbf M}}

\def\I{{\mathbf I}}
\def\Tr{{\rm Tr}}
\def\R{\mathbb{R}}

\title{Neuroscience-inspired online unsupervised learning algorithms}
\date{}

\author[1]{Cengiz Pehlevan\thanks{cpehlevan@seas.harvard.edu}}
\author[2,3]{Dmitri B. Chklovskii\thanks{dchklovskii@flatironinstitute.org}}
\affil[1]{John A. Paulson School of Engineering and Applied Sciences, Harvard University}
\affil[2]{Flatiron Institute, Simons Foundation }
\affil[3]{Neuroscience Institute, NYU Medical Center}

\begin{document}
\maketitle
\doublespacing
\begin{abstract}
Although the currently popular deep learning networks achieve unprecedented performance on some tasks, the human brain still has a monopoly on general intelligence. Motivated by this and  biological implausibility of deep learning networks, we developed a family of biologically plausible artificial neural networks (NNs) for unsupervised learning. Our approach is based on optimizing principled objective functions containing a term that matches the pairwise similarity of outputs to the similarity of inputs, hence the name - similarity-based. Gradient-based online optimization of such similarity-based objective functions can be implemented by NNs with  biologically plausible local learning rules. Similarity-based cost functions and associated NNs solve unsupervised learning tasks such as linear  dimensionality  reduction,  sparse  and/or  nonnegative  feature  extraction, blind nonnegative source separation, clustering and manifold learning.  

\end{abstract}

\section{Introduction}

Inventors of the original artificial neural networks (NNs) derived their inspiration from biology \cite{rosenblatt1958perceptron}. However, today, most artificial NNs such as, for example, backpropagation-based  convolutional  deep  learning networks, resemble natural NNs only superficially. Given that, on some tasks, such artificial NNs achieve human or even superhuman performance, why should one care about such dissimilarity with natural NNs? The algorithms of natural NNs are relevant if one's goal is not just to outperform humans on certain tasks but to develop general-purpose artificial intelligence rivaling human. As contemporary artificial NNs are far from achieving this goal and natural NNs, by definition, achieve it, natural NNs must contain some “secret sauce” that artificial NNs lack. This is why we need to understand the algorithms implemented by natural NNs.
 
Motivated by this argument, we have been developing artificial NNs that could plausibly model natural NNs on the algorithmic level. In our artificial NNs, we do not attempt to reproduce many biological details, as in existing biophysical modeling work, but rather develop algorithms that respect major biological constraints. 
 
For example, biologically plausible algorithms must be formulated in the {\bf online} (or streaming), rather than offline (or batch), setting. This means that input data are streamed to the algorithm sequentially, and the corresponding output must be computed before the next input sample arrives. Moreover, memory accessible to a biological algorithm is limited so that no significant fraction of previous input or output can be stored. 
 
Another key constraint is that in biologically plausible NNs, learning rules must be {\bf local}: a biological synapse can update its weight based on the activity of only the two neurons that the synapse connects. Such “locality” of the learning rule is violated by most artificial NNs including backpropagation-based deep learning networks. In contrast, our NNs employ exclusively local learning rules. Such rules are also helpful for hardware implementations of artificial NNs in neuromorphic chips \cite{davies2018loihi,poikonen2017mixed}. 
 
We derive the algorithms performed by our NNs from optimization objectives. In addition to deriving  learning rules for synaptic weights, as is done in existing artificial NNs, we also derive the architecture, activation functions, and dynamics of neural activity from the same objectives. To do this, we postulate only a cost function and an optimization algorithm, which in our case is alternating stochastic gradient descent-ascent \cite{olshausen1996emergence}. The steps of this algorithm map to a NN, specifying its architecture, activation functions, dynamics, and learning rules. Viewing both weight and activity updates as the steps of an online optimization algorithm allows us to predict the output of our NNs to a wide range of stimuli without relying on exhaustive numerical simulation.     
 
To derive local learning rules we employ optimization objectives operating with pairwise similarities of inputs and outputs of a NN rather than individual data points. Typically, our objectives favor similar outputs for similar inputs. Hence, the name - similarity matching objectives. The transformation of dissimilar inputs in the NN depends on the optimization constraints. Despite using pairwise similarities we still manage to derive {\it online} optimization algorithms.

Our focus is on {\bf unsupervised} learning. This is not a hard constraint, but rather a matter of priority. Whereas humans are clearly capable of supervised learning, most of our learning tasks lack big labeled datasets. On the mechanistic level, most neurons lack a clear supervision signal.

This paper is organized as follows. We start by presenting the conventional approach to deriving unsupervised NNs (Section 2). While the conventional approach generates a reasonable algorithmic model of a single biological neuron, multi-neuron networks violate biological constraints. To overcome this difficulty, in Section 3, we introduce similarity-based cost functions and show that linear dimensionality reduction NNs derived from such cost functions are biologically plausible. In Section 4, we introduce a sign-constrained similarity-matching objective and discuss algorithms for sparse feature extraction and nonnegative independent component analysis. In Section 5, we discuss other sign-constrained networks for clustering and manifold learning. We conclude by discussing potential applications of our work to neuromorphic computing and charting future directions.

\section{Background}

\subsection{Single-neuron online Principal Component Analysis (PCA)}

In the seminal 1982 paper \cite{oja1982simplified}, Oja proposed that a biological neuron can be viewed as an implementation of a mathematical algorithm solving a computational objective. He proposed to model a neuron by an online Principal Component Analysis (PCA) algorithm. As PCA is a workhorse of data analysis used for dimensionality reduction, denoising, and latent factor discovery, Oja's model offers an algorithmic-level description of biological NNs.

Oja’s single-neuron online PCA algorithm works as follows. At each time step, $t$, it receives an input data sample, $\x_t\in\R^n$. As our focus is on the online setting, we use the same variable, $t$, to measure time and index the data points. Then, the algorithm computes and outputs the corresponding top principal component value, $y_t\in\R$:
\begin{align}\label{oja_neuron}
 y_t \longleftarrow \w_{t-1}^\top\x_t,  
\end{align}
where $\w_{t-1}\in\R^n$ is the feature vector computed at time step, $t-1$.  Here and below lowercase italic letters are scalar variables and lowercase boldfaced letters designate vectors. 

At the same time step, $t$, after computing the principal component, the algorithm updates the (normalized) feature vector with a learning rate, $\eta$,
\begin{align}\label{oja_update}
\w_t\longleftarrow \w_{t-1}+\eta\left(\x_t-\w_{t-1} y_t\right)y_t.
\end{align}
If data are drawn i.i.d. from a stationary distribution with a mean vector of zero, the feature vector, $\w$, converges to the eigenvector corresponding to the largest eigenvalue of input covariance \cite{oja1982simplified}.

The steps of the Oja algorithm \eqref{oja_neuron},\eqref{oja_update} naturally correspond to the operations of a biological neuron. Assuming that the components of the input vector are represented by the activities (firing rates) of the upstream neurons, \eqref{oja_neuron} describes a weighted summation of the inputs by the output neuron. Such weighted summation can be naturally implemented by storing the components of feature vector, $\bf w$, in the corresponding synaptic weights. If the activation function is linear, the output, $y_t$, is simply the weighted sum. 

The weight update \eqref{oja_update} is a biologically plausible local synaptic learning rule. The first term of the update, $\x_t y_t$, is proportional to the correlation of the pre- and postsynaptic neurons' activities and the second term, $\w_t y_t^2$, also local, asymptotically normalizes the synaptic weight vector to one. In neuroscience, synaptic weight updates proportional to the correlation of the pre- and postsynaptic neurons' activities are called Hebbian.

\subsection{Minimization of the reconstruction error yields biologically implausible multi-neuron networks}

Next, we would like to build on Oja’s insightful identification of biological processes with the steps of the online PCA algorithms by computing multiple principal components using multi-neuron NNs. Instead of trying to extend the Oja model heuristically, we will derive them by using optimization of a principled objective function. Specifically, we postulate that the algorithm minimizes the reconstruction error, derive an online algorithm optimizing such objective, and map the steps of the algorithm onto biological processes.

In the conventional reconstruction error minimization approach, each data sample, $\x_t\in\R^n$, is approximated as a linear combination of each neuron's feature vector weighted by its activity \cite{olshausen1996emergence}. Then the minimization of the reconstruction (or coding) error can be expressed as follows:
\begin{align}\label{rE}
    \min_{\W } \sum_{t=1}^T \min_{\y_t}\left\Vert \x_t -\W \y_t\right\Vert_2^2,
\end{align}
where matrix $\W\in\R^{n\times k}$, $k<n$, is a concatenation of feature column-vectors and $T$ is both the number of data samples and (in the online setting) the number of time steps.

In the offline setting, a solution to the optimization problem \eqref{rE} is PCA: the columns of optimum $\W$ are a basis for the $k$-dimensional principal subspace \cite{udell2016generalized}. Elements of $\W$ could be constrained to avoid unreasonably low or high values.

In the online setting, \eqref{rE} can be solved by alternating minimization \cite{olshausen1996emergence}. After the arrival of data sample, $\x_t$, the feature vectors are kept fixed while the objective \eqref{rE} is minimized with respect to the principal components by running the following gradient-descent dynamics until convergence:
\begin{align}\label{oja_multineuron}
    \dot \y_t = \W^\top_{t-1}\x_t - \W^\top_{t-1}\W_{t-1}\y_t,
\end{align}
where $\dot{}$ is a derivative with respect to a continuous time variable which runs within a time step, $t$. Unlike a closed-form output of a single Oja neuron \eqref{oja_neuron}, \eqref{oja_multineuron} is iterative. 

After the output, ${\bf y}_t$ converges, at the same time step, $t$, the feature vectors are updated according to the following gradient-descent step, with respect to $\bf W$ on the total objective:
\begin{align}\label{oja_multiupdate}
    \W_t \longleftarrow \W_{t-1} + \eta \left(\x_t - \W_{t-1}\y_t\right)\y_t^\top.
\end{align}
If there was a single output channel, the algorithm \eqref{oja_multineuron},\eqref{oja_multiupdate} would reduce to \eqref{oja_neuron},\eqref{oja_update}, provided that the scalar $\W^\top_{t-1}\W_{t-1}$ is rescaled to unity.

\begin{figure}[t]
\centering
\includegraphics[width = \textwidth]{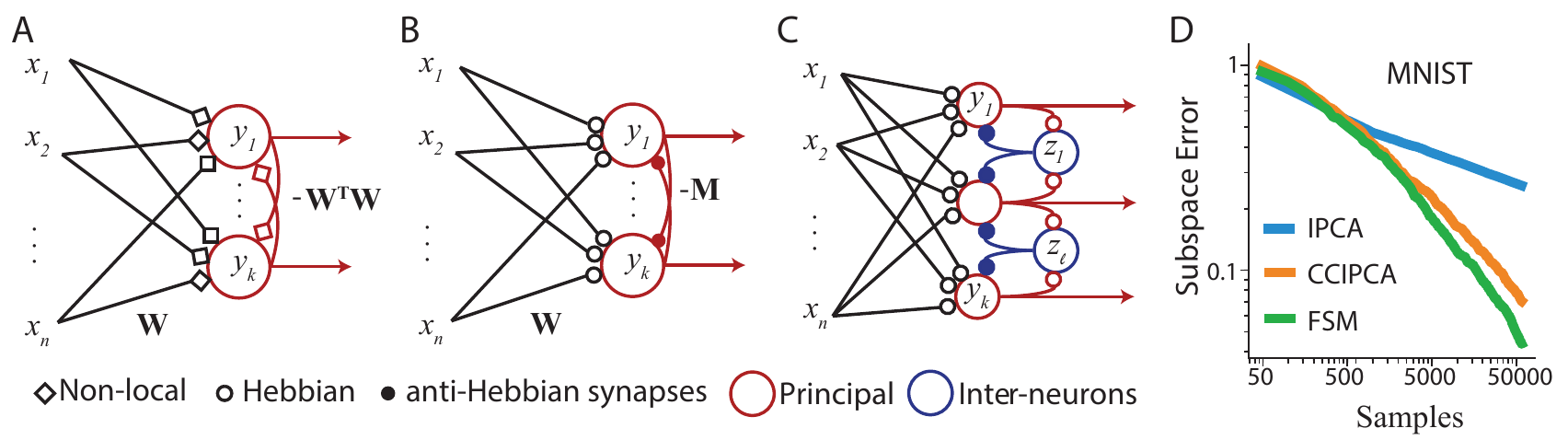}
\caption{\label{fig:SMnet} A) The single-layer NN implementation of the multi-neuron online PCA algorithm derived using the reconstruction approach requires nonlocal learning rules. B) A Hebbian/Anti-Hebbian network derived from similarity matching. C) A biologically-plausible NN with multiple populations of neurons for whitening inputs, derived from a constrained similarity-alignment cost function \cite{pehlevan2015normative}. D) Performance of the Fast Similarity Matching (FSM) algorithm, compared to state-of-the-art online PCA algorithms \cite{arora2012stochastic,weng2003candid} for the MNIST dataset reduced to $k=16$ dimensions (See \cite{giovannucci2018efficient} for details and error definitions). }
\end{figure}

In NN implementations of algorithm \eqref{oja_multineuron},\eqref{oja_multiupdate}, feature vectors are represented by synaptic weights and components by the activities of output neurons. Then \eqref{oja_multineuron} can be implemented by a single-layer NN, Fig. \ref{fig:SMnet}A, in which activity dynamics converges faster than the time interval between the arrival of successive data samples. The lateral connection weights, $-\W^\top_{t-1}\W_{t-1}$, decorrelate neuronal feature vectors by suppressing activities of correlated neurons.

However, implementing update \eqref{oja_multiupdate} in the single-layer NN architecture, Fig. \ref{fig:SMnet}A, requires nonlocal learning rules making it biologically implausible. Indeed, the last term in \eqref{oja_multiupdate} implies that updating the weight of a synapse requires the knowledge of output activities of all other neurons which are not available to the synapse. Moreover, the matrix of lateral connection weights, $-\W^\top_{t-1}\W_{t-1}$, in the last term of \eqref{oja_multineuron} is computed as a Grammian of feedforward weights, clearly a nonlocal operation. This problem is not limited to PCA and arises in networks of nonlinear neurons as well \cite{olshausen1996emergence,lee1999learning}.

To respect the local learning constraint, many authors constructed biologically plausible single-layer networks using heuristic local learning rules that were not derived from an objective function \cite{foldiak1989adaptive,diamantaras1996principal}. However, we think that abandoning the optimization approach creates more problems than it solves. Alternatively, NNs with local learning rules can be derived if one introduces a second layer of neurons  \cite{olshausen1997sparse}. However, such architecture does not map naturally on biological networks.

We have outlined how the conventional reconstruction approach fails to generate biologically plausible multi-neuron networks for online PCA. In the next section, we will introduce an alternative approach that overcomes this limitation. Moreover, this approach suggests a novel view of neural computation leading to many interesting extensions.

\section{Similarity-based approach to linear dimensionality reduction}

In this section, we propose a different objective function for the optimization approach to constructing PCA NNs, which we term similarity matching. From this objective function, we derive an online algorithm implementable by a NN with local learning rules. Then, we introduce other similarity-based algorithms for linear dimensionality reduction which include more biological features such as different neuron classes.

\subsection{Similarity-matching objective function}

We start by stating an objective function that will be used to derive NNs for linear dimensionality reduction. The similarity of a pair of inputs, $\x_t$ and $\x_{t'}$, both in $\R^n$, can be defined as their dot-product, $\x_t^\top \x_{t'} $. Analogously, the similarity of a pair of outputs, which live in $\R^k$ with $k<n$, is $\y_{t}^\top \y_{t'}$. Similarity matching, as its name suggests, learns a representation where the similarity between each pair of outputs matches that of the corresponding inputs:
\begin{align}\label{CMDS}
\min_{ \y_1,\ldots,\y_T}  \frac{1}{T^2}  \sum_{t=1}^T \sum_{t'=1}^T \left(\x_t^\top \x_{t'} - \y_t^\top\y_{t'}\right)^2.
\end{align}
This offline objective function, previously employed for multidimensional scaling, is optimized by the projections of inputs onto the principal subspace of their covariance, i.e. performing PCA up to an orthogonal rotation. Moreover, \eqref{CMDS} has no local minima other than the principal subspace solution \cite{pehlevan2015normative,GeLeeMa}.  

The similarity-matching objective \eqref{CMDS} may seem like a strange choice for deriving an online algorithm implementable by a NN. In \eqref{CMDS}, pairs of inputs and outputs from different time steps interact with each other. Yet, in the online setting, an output must be computed at each time step without accessing inputs or outputs from other time steps. In addition, synaptic weights do not appear explicitly in \eqref{CMDS} seemingly precluding mapping onto a NN. 

\subsection{Variable substitution trick}

Both of the above concerns can be resolved by a simple math trick akin to completing the square \cite{pehlevan2018similarity}. We first focus on the cross-term in the expansion of the square in \eqref{CMDS}, which we call similarity alignment. By introducing a new variable, ${\bf W} \in  \R^{k\times n}$, we can re-write the cross-term:
\begin{align}\label{cross}
 &- \frac{1}{T^2}\sum_{t=1}^T \sum_{t'=1}^T  \y_{t}^\top \y_{t'} \x_t^\top \x_{t'} =   \min_{\W\in \R^{k\times n}} \, -\frac{2}{T} \sum_{t=1}^T\y_t^\top \W \x_t + \Tr \, \W^\top \W.
\end{align}
To prove this identity, find optimal ${\bf W}$ by taking a derivative of the expression on the right with respect to ${\bf W}$ and setting it to zero, and then substitute the optimal ${\bf W}^* = \frac 1T\sum_{t=1}^T \y_t\x_t^\top$ back into the expression. Similarly, for the quartic $\y_t$ term in \eqref{CMDS}:
\begin{align}\label{quartic}
 &\frac 1{T^2}\sum_{t=1}^T \sum_{t'=1}^T  \y_{t}^\top \y_{t'} \y_t^\top \y_{t'}  =   \max_{\M \in \R^{k\times k}} \, \frac{2}{T} \sum_{t=1}^T\y_t^\top \M \y_t - \Tr \, \M^\top \M.
\end{align}
By substituting \eqref{cross} and \eqref{quartic} into \eqref{CMDS} we get:
\begin{align}\label{SMMW2t}
\min_{{\bf W}\in \mathbb{R}^{k\times n}}\max_{{\bf M}\in \mathbb{R}^{k\times k}}  \, \frac{1}{T} \sum_{t=1}^T \left[2\, {\rm Tr}\left({\bf W}^\top{\bf W}\right) - {\rm Tr}\left({\bf M}^\top{\bf M}\right) +  \min_{{\bf y}_t\in \mathbb{R}^{k\times 1}}  l_t({\bf W},{\bf M},{\bf y}_t)\right],
\end{align}  
where
\begin{align}\label{lyapunov}
l_t({\bf W},{\bf M},{\bf y}_t)=-4{\bf x}_t^\top{\bf W}^\top{\bf y}_t + 2{\bf y}_t^\top{\bf M}{\bf y}_t.
\end{align} 
In the resulting objective function, \eqref{SMMW2t},\eqref{lyapunov}, optimal outputs at different time steps can be computed independently, making the problem amenable to an online algorithm. The price paid for this simplification is the appearance of the minimax optimization problem in variables, {\bf W} and {\bf M}. Minimization over $\bf W$ aligns output channels with the greatest variance directions of the input and maximization over $\bf M$ diversifies the output by decorrelating output channels similarly to the Grammian, $\W^\top\W$, used previously. This competition in a gradient descent/ascent algorithm results in the principal subspace projection which is the only stable fixed point of the corresponding dynamics \cite{pehlevan2015MDS}.

\subsection{Online algorithm and neural network}

We are ready to derive an algorithm for optimizing \eqref{CMDS} online. First, we  minimize \eqref{lyapunov} with respect to the output variables, ${\bf y}_t$, while holding ${\bf W}$ and ${\bf M}$ fixed using gradient-descent dynamics:
\begin{align}
\label{grad}
    \dot\y_t=\W \x_t-\M \y_t.
\end{align}
As before, dynamics \eqref{grad} converges within a single time step, $t$, and outputs $\y_t$. After the convergence of $\y_t$, we update ${\bf W}$ and ${\bf M}$ by gradient descent of \eqref{cross} and gradient ascent of \eqref{quartic} respectively:
\begin{align}
\label{Hebb}
W_{ij} \leftarrow W_{ij} + \eta \left(y_ix_j-W_{ij}\right), \qquad M_{ij} \leftarrow M_{ij} + \frac{\eta}2\left(y_iy_j-M_{ij}\right).
\end{align}

Algorithm \eqref{grad},\eqref{Hebb}, first derived in \cite{pehlevan2015MDS}, can be naturally implemented by a biologically plausible NN, Fig. \ref{fig:SMnet}B. Here, activity of the upstream neurons corresponds to input variables. Output variables are computed by the dynamics of activity \eqref{grad} in a single layer of neurons. Variables ${\bf W}$ and ${\bf M}$ are represented by the weights of synapses in feedforward and lateral connections respectively. The learning rules \eqref{Hebb} are local, i.e. the weight update, $\Delta W_{ij}$, for the synapse between $j^{\rm th}$ input neuron and $i^{\rm th}$ output neuron depends only on the activities, $x_j$, of $j^{\rm th}$ input neuron and, $y_i$, of $i^{\rm th}$ output neuron, and the synaptic weight. In neuroscience, learning rules \eqref{Hebb} for synaptic weights ${\bf W}$ and ${\bf -M}$ (here minus indicates inhibitory synapses, see Eq.\eqref{grad}) are called Hebbian and anti-Hebbian respectively.

To summarize this Section so far, starting with the similarity-matching objective, we derived a Hebbian/anti-Hebbian NN for dimensionality reduction. The minimax objective  can be viewed as a zero-sum game played by the weights of feedforward and lateral connections \cite{pehlevan2018similarity,seung2017correlation}. This demonstrates that synapses with local updates can still collectively work together to optimize a global objective. A similar, although not identical, NN was proposed by F\"oldiak \cite{foldiak1989adaptive} heuristically. The advantage of our optimization approach is that the offline solution is known. 

Although no proof of convergence exists in the online setting, algorithm \eqref{grad},\eqref{Hebb} performs well on large-scale data. A recent paper \cite{giovannucci2018efficient}  introduced an efficient, albeit non-biological, modification of the similarity-matching algorithm, Fast Similarity Matching (FSM), and demonstrated its competitiveness with the state-of-the-art principal subspace projection algorithms in both processing speed and convergence rate\footnote{A package with implementations of these algorithms is on \href{https://github.com/flatironinstitute/online_psp}{https://github.com/flatironinstitute/online\_psp} and \href{https://github.com/flatironinstitute/online_psp_matlab}{https://github.com/flatironinstitute/online\_psp\_matlab}.}, Figure \ref{fig:SMnet}D.  FSM produces the same output $\y_t$ for each input $\x_t$ as similarity-matching by optimizing \eqref{lyapunov} by matrix inversion, $\y_t = \M^{-1}\W\x_t$. It achieves extra computational efficiency by keeping in memory and updating the $\M^{-1}$ matrix rather than $\M$. We refer the reader to \cite{giovannucci2018efficient} for suggestions on the implementation of these algorithms.

\subsection{Other similarity-based objectives and linear networks}

 As the algorithm \eqref{grad},\eqref{Hebb} and the NN in Fig.\ref{fig:SMnet}B were derived from the similarity-matching objective \eqref{CMDS}, they project data onto the principal subspace but do not necessarily recover principal components {\it per se}. To derive PCA algorithms we modified the objective function \eqref{CMDS} to encourage orthogonality of $\W$ \cite{pehlevan2015optimization,minden2018biologically}. Such algorithms are implemented by NNs of the same architecture as in Fig.\ref{fig:SMnet}B but with slightly different local learning rules.  

Although the similarity-matching NN in Fig. \ref{fig:SMnet}B relies on biologically plausible local learning rules, it lacks biological realism in several other ways. For example, computing output requires recurrent activity that must settle faster than the time scale of the input variation, which is unlikely in biology. To respect this biological constraint, we modified the dimensionality reduction algorithm to avoid recurrency \cite{minden2018biologically}.

Another non-biological feature of the NN in Fig.\ref{fig:SMnet}B is that the output neurons compete with each other by communicating via lateral connections. In biology, such interactions are not direct but mediated by interneurons. To reflect these observations, we modified the objective function by introducing a whitening constraint: 
\begin{align}\label{NIPS3}
&\min_{\y_1,\ldots,\y_T}  - \frac{1}{T^2}\sum_{t=1}^T \sum_{t'=1}^T  \y_{t}^\top \y_{t'} \x_t^\top \x_{t'}, \qquad {\rm s.t.} \quad \frac 1T \sum_{t}\y_t\y_t^\top = \I_k,
\end{align}
where $\I_k$ is the $k$-by-$k$ identity matrix. Then, by implementing the whitening constraint using the Lagrange formalism, we derived NNs where interneurons appear naturally - their activity is modeled by the Lagrange multipliers, $\z_t^\top \z_{t'}$ (Fig. \ref{fig:SMnet}C), \cite{pehlevan2015normative}:
\begin{align}\label{NIPS3_minmax}
\min_{\y_1,\ldots,\y_t} \max_{\z_1,\ldots,\z_T } &- \frac{1}{T^2}\sum_{t=1}^T \sum_{t'=1}^T  \y_{t}^\top \y_{t'} \x_t^\top \x_{t'} + \frac{1}{T^2}\sum_{t=1}^T \sum_{t'=1}^T  \z_{t}^\top \z_{t'} \left(\y_{t}^\top \y_{t'} -\delta_{t,t'}\right),
\end{align}
where $\delta_{t,t'}$ is the Kronecker delta. Notice how \eqref{NIPS3_minmax} contains the $\y$-$\z$ similarity-alignment term similar to \eqref{cross}. We can now derive learning rules for the $\y$-$\z$ connections using the variable substitution trick, leading to the network in Figure \ref{fig:SMnet}C. For details of this and other NN derivations, see \cite{pehlevan2015normative}.

\section{Nonnegative similarity-matching objective and nonnegative independent component analysis}

\begin{figure}
\begin{floatrow}
\ffigbox[\FBwidth]{%
  \includegraphics[scale = 0.8]{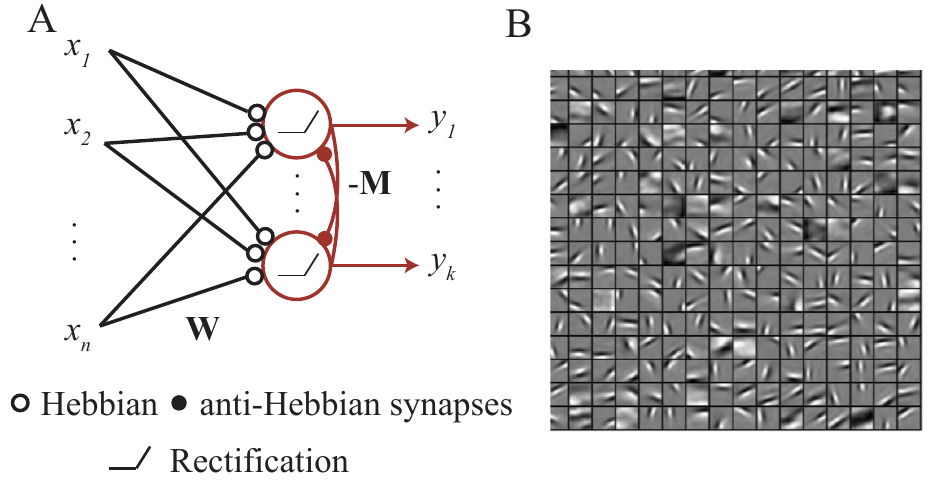}
}{%
  \caption{\label{fig:Gabor} A) A nonlinear Hebbian/Anti-Hebbian network derived from nonnegative similarity matching. B) Nonnegative similarity matching learns edge filters from patches of whitened natural scenes. Learned filters in small squares. See \cite{pehlevan2014NMF} for details of the simulations.}
}
\capbtabbox{%

\begin{tabular}{ |p{4cm}|p{1.55cm}|  }
\hline
Algorithm & Accuracy \\
\hline
\hline
\textbf{Convolutional Nonnegative Similarity Matching} \cite{bahroun2017online}   & \textbf{80.42 \%} \\
\hline
\hline
K-means \cite{coates2011analysis}  & 79.60 \%  \\
Convolutional DBN \cite{krizhevsky2010convolutional} & 78.90 \%  \\
\hline
\end{tabular}
 
}{%
  \caption{Performance of unsupervised feature learning algorithms. We list linear classification accuracy on CIFAR-10 using features extracted by nonnegative similarity matching network. We also list the best single-layer feature extractor (K-means) from an earlier study \cite{coates2011analysis} and a deep belief network \cite{krizhevsky2010convolutional} on the same task. Detailed comparisons are available in \cite{bahroun2017online}.}\label{Tab:Multi_resolution}
}
\end{floatrow}
\end{figure}

So far we considered  similarity-based NNs comprising linear neurons. But many interesting computations require nonlinearity and biological neurons are not linear. To construct more realistic and powerful similarity-based NNs, we note that the output of biological neurons is nonnegative (firing rate cannot be below zero). Hence, we modified the optimization problem by requiring that the output of the similarity-matching cost function \eqref{CMDS} is nonnegative:
\begin{align}\label{nonnegative}
\min_{ \y_1,\ldots,\y_T\ge 0}  \frac{1}{T^2}  \sum_{t=1}^T \sum_{t'=1}^T \left(\x_t^\top \x_{t'} - \y_t^\top\y_{t'}\right)^2.
\end{align}
Here, the number of output dimensions, $k$, may be greater than the number of input dimensions, $n$, leading to a dimensionally expanded representation. Eq. \eqref{nonnegative} can be solved by the same online algorithm as  \eqref{CMDS} except that the output variables are projected onto the nonnegative domain. Such algorithm maps onto the same network and same learning rules as in Eq.\eqref{Hebb}, Fig. \ref{fig:SMnet}B, but with rectifying neurons (ReLUs) \cite{pehlevan2014NMF,pehlevan2017blind,pehlevan2019spiking}, Fig. \ref{fig:Gabor}A. 

Nonnegative similarity-matching network learns features that are very different from PCA. For example, when the network is trained on whitened natural scenes it extracts edge filters \cite{pehlevan2014NMF} (Fig. \ref{fig:Gabor}) as opposed to Fourier harmonics expected for a translationally invariant dataset. Motivated by this observation, Bahroun and Soltoggio \cite{bahroun2017online} developed a convolutional nonnegative similarity matching network with multiple resolutions, and used it as an unsupervised feature extractor for subsequent linear classification on CIFAR-10 dataset. They found that nonnegative similarity matching NNs are superior to other single-layer unsupervised techniques \cite{bahroun2017online,bahroun2017building}, Table \ref{Tab:Multi_resolution}. 

\begin{figure}[tb]
\centering
\includegraphics[scale = 0.95]{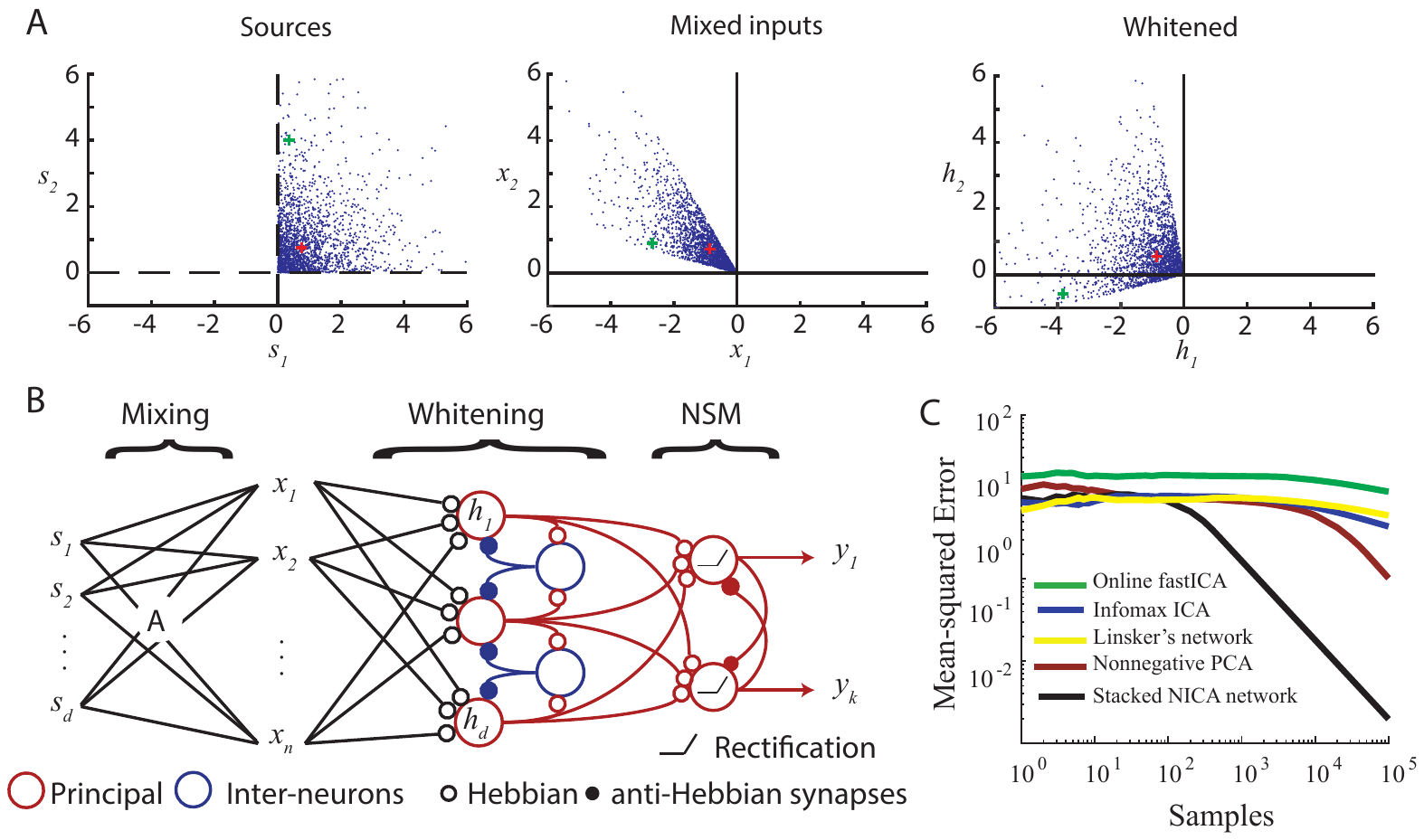}
\caption{\label{fig:NICA}  A) Illustration of Plumbley's nonnegative independent component analysis algorithm. Two source channels (left, each blue point shows one source vector, $\s_t$) are
linearly mixed to a two-dimensional mixture, which are the inputs to the algorithm (middle). Whitening (right)
yields an orthogonal rotation of the sources. Sources are then recovered by
solving the nonnegative similarity-matching problem. Green and red plus signs track two source vectors
across mixing and whitening stages. B) Stacked network for NICA. The network sees the mixtures $\x_t$ and aims to recover sources $\s_t$ at its output $\y_t$. NSM - nonnegative similarity-matching. C) Performance of the stacked network for NICA (black) in reconstructing the source vector on a 10-dimensional artificial dataset, see \cite{pehlevan2017blind} for details. Performance metric is the squared error between the network's output and the original sources, averaged over all samples until that point.}
\end{figure}

As edge filters emerge also in the independent component analysis (ICA) of natural scenes  \cite{bell1997independent} we investigated a connection of nonnegative similarity matching with nonnegative independent component analysis (NICA) used for blind source separation. The NICA problem is to recover independent, nonnegative and well-grounded (finite probability
density function in any positive neighborhood of zero) sources, $\s_t\in\R^d$, from  observing only their linear mixture, $\x_t={\bf A}\s_t$, where ${\bf A}\in \R^{n\times d}$, and $n\geq d$. 

Our solution of NICA is based on the observation that NICA can be solved in two steps \cite{plumbley2002conditions}, Fig. \ref{fig:NICA}A. First, whiten the data and reduce it to $d$ dimensions to obtain an orthogonal rotation of the sources (assuming that the mixing matrix is full-rank). Second, find an orthogonal rotation of the whitened sources that yields a nonnegative output, Fig. \ref{fig:NICA}A.  The first step can be implemented by the whitening network in Fig. \ref{fig:SMnet}C. The second step can be implemented by the nonnegative similarity-matching network (Fig. \ref{fig:Gabor}A) because an orthogonal rotation does not affect dot-product similarities \cite{pehlevan2017blind}. Therefore, NICA is solved by stacking the whitening and the nonnegative similarity-matching networks,  Fig. \ref{fig:NICA}B. This algorithm performs well compared to other popular NICA algorithms \cite{pehlevan2017blind}, Fig. \ref{fig:NICA}C.

\section{Non-negative similarity-based networks for clustering and manifold tiling}

Nonnegative similarity-matching can also cluster well-segregated data \cite{kuang2012symmetric,pehlevan2014NMF} and, for data concentrated on manifolds, it can tile them \cite{sengupta2018manifold}. To understand this behavior, we analyze the optimal solutions of nonnegative similarity-based objectives. Finding the optimal solution for a constrained similarity-based objective is rather challenging as has been observed for the non-negative matrix factorization problem. Here, we introduce a simplified similarity-based objective that allows us to make progress with the analysis and admits an intuitive interpretation. First, we address the simpler clustering task which, for highly segregated data, has a straightforward optimal solution. Second, we address manifold learning by viewing it as a soft-clustering problem. 

\subsection{A similarity-based cost function and NN for clustering}

The key to our analysis is formulating a similarity-based cost function, an optimization of which will yield an online algorithm and a NN for clustering. The algorithm should assign inputs $\x_t$ to $k$ clusters based on pairwise similarities and output cluster assignment indices $\y_{t}$. 

To arrive at a cost function, consider first a single pair of data points, $\x_1$ and $\x_2$. If their similarity, $\x_1^\top \x_{2}<\alpha$, where $\alpha$ is a pre-set threshold, then the points should be assigned to separate clusters, i.e. $\y_1 = [1,0]^\top$ and  $\y_2 = [0,1]^\top$, setting output similarity, ${\bf y}_1^\top{\bf y}_{2}$ to 0. If $\x_1^\top \x_{2}>\alpha$, then the points are assigned to the same cluster, e.g. ${\bf y}_1={\bf y}_{2} = [1,0]^\top$. Such $\y_1$ and $\y_2$ are optimal solutions (although not unique) to the following optimization problem:
\begin{align}\label{Eq_single}
\min_{{\bf y}_1\geq 0,{\bf y}_2\geq 0} \left(\alpha-\x_1^\top \x_2\right){\bf y}_1^\top{\bf y}_{2}, \qquad {\rm s.t.} \quad \left\Vert{\bf y}_1\right\Vert_2\leq 1, \, \left\Vert{\bf y}_{2}\right\Vert_2\leq 1.
\end{align}

To obtain an objective function that would cluster the whole dataset of $T$ inputs we simply sum \eqref{Eq_single} over all possible input pairs:
\begin{align}\label{Eq_obj}
\min_{{\bf y}_1\geq 0,\ldots,{\bf y}_T\geq 0} \sum_{t=1}^T\sum_{t'=1}^T\left(\alpha-\x_t^\top \x_{t'}\right){\bf y}_t^\top{\bf y}_{t'} \qquad 
{\rm s.t.}\quad \left\Vert{\bf y}_1\right\Vert_2\leq 1, \quad\ldots\quad ,\left\Vert{\bf y}_T\right\Vert_2\leq 1.
\end{align}

Does optimization of \eqref{Eq_obj} produce the desired clustering output? This depends on the dataset. If a threshold, $\alpha$, exists such that the similarities of all pairs within the same cluster are greater and similarities of pairs from different clusters are less than $\alpha$, then the cost function \eqref{Eq_obj} is minimized by the desired hard-clustering output, provided that $k$ is greater than or equal to the number of clusters. 

\begin{figure}[t]
\centering
\includegraphics[width=1\textwidth]{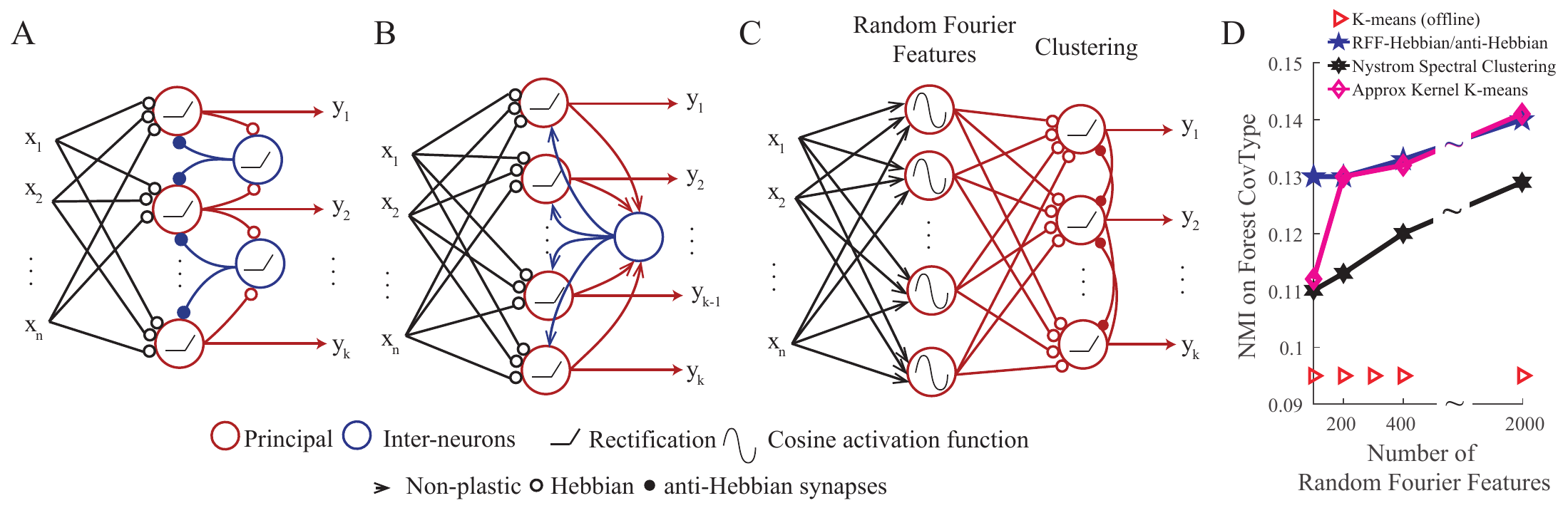}
\caption{\label{fig:MNets}  Biologically-plausible NNs for clustering and manifold learning. A) A biologically-plausible excitatory-inhibitory NN implementation of the algorithm. In this version, anti-Hebbian synapses operate at a faster time scale than Hebbian synapses \cite{sengupta2018manifold}. B) Hard and soft $k$-means networks. Rectified neurons are perfect (hard $k$-means) or leaky (soft $k$-means) integrators. They have learned (homeostatic) activation thresholds and ephaptic couplings. C) When augmented with a hidden nonlinear layer, the presented networks perform clustering in the nonlinear feature space. Shown is the NN of \cite{bahroun2017neural}, where the hidden layer is formed of Random Fourier Features \cite{rahimi2008random} to obtain a low-rank approximation to a Gaussian kernel. The two-layer NN operates as an online kernel clustering algorithm, and D) performs on par to other state-of-the-art kernel clustering algorithms \cite{bahroun2017neural}. Shown is performance (Normalized Mutual Information - NMI) on Forest Cover Type dataset. Figure modified from \cite{bahroun2017neural}.}
\end{figure}

To solve the objective \eqref{Eq_obj} in the online setting, we introduce the constraints in the cost via Lagrange multipliers and using the variable substitution trick, we can derive a NN implementation of this algorithm \cite{sengupta2018manifold} (Fig. \ref{fig:MNets}A). The algorithm operates with local Hebbian and anti-Hebbian learning rules, whose functional form is equivalent to \eqref{Hebb}.

\subsection{Manifold-tiling solutions}

In many real-world problems, data points are not well-segregated but lie on low-dimensional manifolds. For such data, the optimal solution of the objective \eqref{Eq_obj} effectively tiles the  data manifold \cite{sengupta2018manifold}.

\begin{figure}[tb]
\centering
\includegraphics[width=\textwidth]{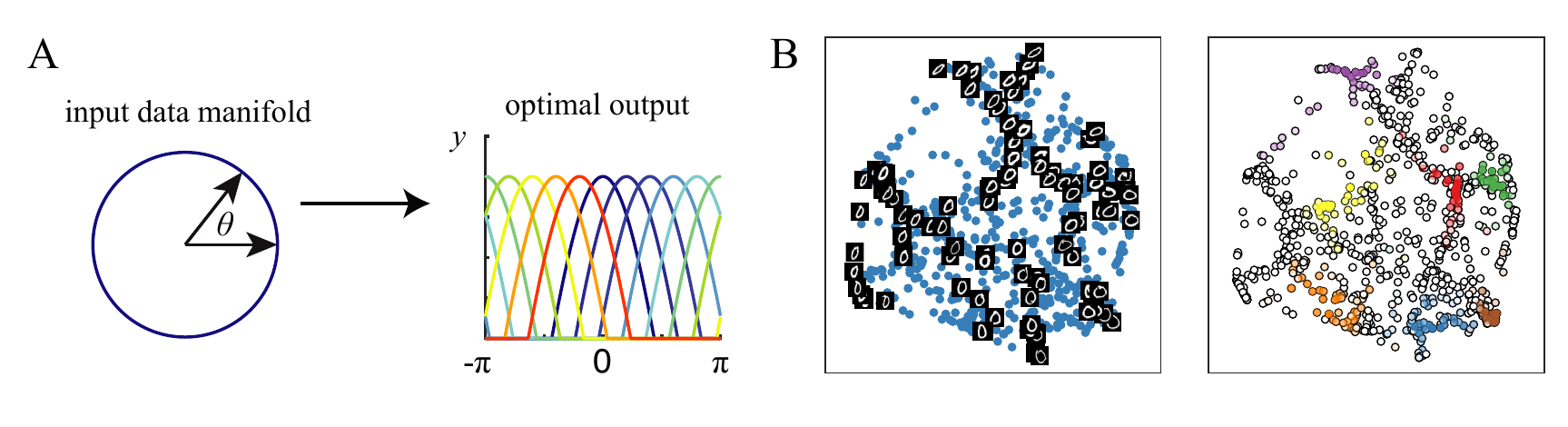}
\caption{\label{fig:tiling} Analytical and numerical manifold-tiling solutions of \eqref{Eq_obj} for representative datasets provide accurate and useful representations. A) A circular manifold (left) is tiled by  overlapping localized receptive fields (right). In the continuum limit ($k\rightarrow \infty$), receptive fields are truncated cosines of the polar angle, $\theta$ \cite{sengupta2018manifold}. Similar analytical and numerical results are obtained for a spherical 3D manifold and SO(3) (not shown, see \cite{sengupta2018manifold}).  B) Learning the manifold of the 0 digit from the MNIST dataset by tiling the manifold with
overlapping localized receptive fields. Left: Two-dimensional
linear embedding (PCA) of the outputs. The data gets organized according to different visual characteristics
of the hand-written digit (e.g., orientation and elongation). Right: Sample receptive fields over the low-dimensional embedding.}
\end{figure}

We can understand such optimal solutions using soft-clustering, i.e. clustering where each stimulus may be assigned to more than one cluster and assignment indices are real numbers between zero and one. Each output neuron is characterized by the weight vector of incoming synapses which defines a centroid in the input data space. The response of a neuron is maximum when data fall on the centroid and decays away from it. Manifold-tiling solutions for several datasets are shown in Fig. \ref{fig:tiling}.

We can prove this result analytically by taking advantage of the convex relaxation in the limit of infinite number of output dimensions, i.e. $k\to\infty$. 
Indeed, if we introduce Gramians ${\bf D}$, such that $D_{tt'}=\x_t^\top \x_{t'}$, and ${\bf Q}$, such that $Q_{tt'}=\y_t^\top \y_{t'}$ and do not specify the dimensionality of $\y$ by leaving the rank of ${\bf Q}$ open,
we can rewrite \eqref{Eq_obj} as: 
\begin{align}
\label{optim}
 \min_{\substack{ {\bf Q}\in \mathcal{CP}^T \\ {\rm diag}{\bf Q} \le{\bf 1} }}
 -\Tr(({\bf D}-\alpha {\bf E}){\bf Q}),
\end{align}
where ${\bf E}$ is a matrix whose elements are all ones, and the cone of {\it completely positive}  $T\times T$ matrices, i.e. matrices ${\bf Q}\equiv{\bf Y}^\top{\bf Y}$ with ${\bf Y}\ge 0$, is denoted by $\mathcal{CP}^T$ \cite{berman2003completely}.  Redefining the variables makes the optimization problem convex. For arbitrary datasets, optimization problems in $\mathcal{CP}^T$ are often intractable for large $T$ \cite{berman2003completely}, despite the convexity. However, for symmetric datasets, i.e. circle, 2-sphere and SO(3), we can optimize \eqref{optim} by analyzing the  Karush–Kuhn–Tucker conditions \cite{sengupta2018manifold} (Fig. \ref{fig:tiling}A).

\subsection{Other similarity-based NNs for clustering and manifold-tiling}

 A related problem to objective \eqref{Eq_obj} is the previously studied convex semidefinite programming relaxation of community detection in graphs \cite{cai2015robust}, which is closely related to clustering. The semidefinite program is related to \eqref{optim} by requiring the nonnegativity of ${\bf Q}$ instead of the nonnegativity of ${\bf Y}$:
\begin{align}
  \min_{ {\bf Q} \succeq 0,\,  {\bf Q}\geq 0, \, {\rm diag}{\bf Q} \le{\bf 1} }
 -\Tr(({\bf D}-\alpha {\bf E}){\bf Q}).
\end{align}

While we chose to present our similarity-based NN approach to clustering and manifold-tiling through the cost function in \eqref{Eq_obj}, similar results can be obtained for other versions of similarity-based clustering objective functions. The nonnegative similarity-matching cost function Eq. \eqref{nonnegative} and the NN derived from it (Fig. \ref{fig:Gabor}A) can be used for clustering and manifold learning as well \cite{kuang2012symmetric,pehlevan2014NMF,sengupta2018manifold}. The $K$-means cost function can be cast into a similarity-based form and a NN (Fig. \ref{fig:MNets}B) can be derived for its online implementation \cite{pehlevan2017clustering}. We introduced a soft-$K$-means cost, also a relaxation of another semidefinite program for clustering \cite{kulis2009semi}, and an associated NN (Fig. \ref{fig:MNets}B) \cite{pehlevan2017clustering}, and showed that they can perform manifold tiling \cite{tepper2017clustering}.

The algorithms we discussed operate with the dot product as a measure of similarity in the inputs. By augmenting the presented NNs by an initial random, nonlinear projection layer (Fig. \ref{fig:MNets}C), it is possible to implement nonlinear similarity measures associated with certain kernels \cite{bahroun2017neural}. A clustering algorithm using this idea is shown to perform on par with other online kernel clustering algorithms \cite{bahroun2017neural}, Fig. \ref{fig:MNets}D.

\section{Discussion}

To overcome the non-locality of the learning rule in NNs derived from the reconstruction error minimization, we proposed a new class of cost functions called similarity-based. To summarize, the first term in the similarity-based cost functions,
\begin{align}\label{sbc}
   \min_{\forall t, \y_t\in \Omega} \left[- \sum_{t=1}^T \sum_{t'=1}^T  \y_{t}^\top \y_{t'} \x_t^\top \x_{t'}+ f(\y_1,...,\y_T)\right],
\end{align}
is the covariance of the similarity of the outputs and the similarity of the inputs.  Hence, the name ``similarity-based" cost functions. Previously, such objectives were used in linear kernel alignment \cite{cristianini2002kernel}. Our key observation is that optimization of objective functions containing such term in the online setting gives rise to local synaptic learning rules \eqref{cross} \cite{pehlevan2018similarity}. 

To derive biologically plausible NNs from \eqref{sbc}, one must choose not just the first term  but also the function, $f$,  and the optimization constraints, $\Omega$, so that the online optimization algorithm is implementable by biological mechanisms. We and others have identified a whole family of such functions and constraints (Table \ref{table:feat}), some of which were reviewed in this article. As a result, we can relate many features of biological NNs to different terms and constraints in similarity-based cost functions and, hence, give them computational interpretations. 

Our framework provides a systematic procedure to design novel NN algorithms by formulating a learning task using similarity-based cost functions. As evidenced by the high-performing algorithms discussed in this paper, our procedure of incorporating biological constraints does not impede but rather facilitates the design process by limiting the algorithm search to a useful part of the NN algorithm space.

\begin{table}[tb]
\caption{Current list of objectives, regularizers and constraints that define a similarity-based optimization problem and solvable by a NN with local learning.}
\centering 
\def\arraystretch{1.2}
\begin{tabular}{l r} 

OPTIMIZATION FEATURE & BIOLOGICAL FEATURE \\
\hline \hline
Similarity (anti-)alignment & (Anti-)Hebbian plasticity \cite{pehlevan2015MDS,pehlevan2018similarity} \\
\hline 
Nonnegativity constraint & ReLU activation function \cite{pehlevan2014NMF,seung2017correlation}  \\ \hline 
Sparsity regularizer & Adaptive neural threshold \cite{hu2014SMF}\\ \hline 
\shortstack{Constrained output correlation matrix} & Adaptive lateral weights \cite{pehlevan2015normative,seung2017correlation}  \\ \hline 
\shortstack{Constrained PSD output Grammian} & Anti-Hebbian  interneurons \cite{pehlevan2015normative} \\ \hline 
Copositive output Grammian  & Anti-Hebbian inhibitory neurons \cite{sengupta2018manifold} \\ \hline 
Constrained activity $l_1$-norm & Giant interneuron \cite{pehlevan2017clustering} \\ \hline 
\hline \\ 
\end{tabular}
\label{table:feat} 
\end{table}

The locality of learning rules in similarity-based NNs makes them naturally suitable for implementation on adaptive neuromorphic systems, which have already been explored in custom analog arrays \cite{poikonen2017mixed}. For broader use in the rapidly growing world of low-power, spike-based hardware with on-chip learning \cite{davies2018loihi}, similarity-based NNs were missing a key ingredient: spiking neurons. Very recent work \cite{pehlevan2019spiking} developed a spiking version of the nonnegative similarity matching network and took a step towards neuromorphic applications.

Despite the successes of similarity-based NNs, many interesting challenges remain. 1) Whereas numerical experiments indicate that our online algorithms perform well, most of them lack  global convergence proofs. Even for PCA networks we can only prove linear stability of the desired solution in the stochastic approximation setting. 2) Motivated by biological learning, which is mostly unsupervised, we focused on unsupervised learning. Yet, supervision, or reinforcement, does take place in the brain. Therefore, it is desirable to extend our framework to supervised, semi-supervised and reinforcement learning settings. Such extensions may be valuable as general purpose machine learning algorithms. 3) Whereas most sensory stimuli are correlated time series, we assumed that data points at different times are independent. How are temporal correlations analyzed by NNs? Solving this problem is important both for modeling brain function and developing general purpose machine learning algorithms. 4) Another challenge is stacking similarity-based NNs. Heuristic approach to stacking yields promising results \cite{bahroun2017online}. Yet, except for the Nonnegative ICA problem introduced in Section 4, we do not have a theoretical understanding of how and why to stack similarity-based NNs. 5) Finally, neurons in biological NNs signal each other using all-or-none spikes, or action potentials, as opposed to real-valued signals we considered. Is there an optimization theory accounting for spiking in biological NNs?

\subsection*{Acknowledgments}
We thank our collaborators Anirvan Sengupta, Mariano Tepper, Andrea Giovannucci, Alex Genkin, Victor Minden, Sreyas Mohan, Yanis Bahroun for their contributions, Andrea Soltoggio for discussions, and Siavash Golkar and Alper Erdogan for commenting on the manuscript. This work was in part supported by a gift from the Intel Corporation.

\end{document}